\documentclass[a4paper,11pt]{article}
\usepackage{pos}

\title{Basis light-front quantization: Advancing a first principles approach for the nucleon}

\author*[a,b]{Chandan~Mondal}
\author[a,b,c]{Siqi~Xu}
\author[a,b]{Yiping~Liu}
\author[a,b]{Jiangshan~Lan}
\author[d,e]{Yang~Li}
\author[a,b]{Xingbo~Zhao}
\author[]{James~P.~Vary$^{c}$~~~({\rm BLFQ~Collaboration})}

\affiliation[a]{Institute of Modern Physics, Chinese Academy of Sciences, Lanzhou, Gansu, 730000, China}

\affiliation[b]{School of Nuclear Physics, University of Chinese Academy of Sciences, Beijing, 100049, China}

\affiliation[c]{Department of Physics and Astronomy, Iowa State University, Ames, IA 50011, USA}

\affiliation[d]{University of Science and Technology of China, Hefei, Anhui 230026, China}

\affiliation[e]{Anhui Center for Fundamental Sciences in Theoretical Physics, University of Science and Technology of China, Hefei 230026, China}

\emailAdd{mondal@impcas.ac.cn}
\emailAdd{xsq234@impcas.ac.cn}
\emailAdd{liuyiping@impcas.ac.cn}
\emailAdd{jiangshanlan@impcas.ac.cn}
\emailAdd{leeyoung1987@ustc.edu.cn}
\emailAdd{xbzhao@impcas.ac.cn}
\emailAdd{jvary@iastate.edu}

\abstract{We present our recent progress in applying the basis light-front quantization approach to investigate the nucleon's structure. We solve its wave functions from the eigenstates of the light-front QCD Hamiltonian using a fully relativistic, nonperturbative approach without an explicit confining potential. These eigenstates are determined for the three-quark, three-quark-gluon, and three-quark-quark-antiquark Fock representations, making them suitable for low-resolution probes. From these wave functions, we compute the nucleon's quark and gluon matter densities, as well as its helicity and transversity distributions, showing qualitative consistency with experimental data. We also determine the quark and gluon helicity contributions to the proton spin and the tensor charges. The resulting light-front wave functions represent a significant step toward a unified description of hadron distribution functions in both longitudinal and transverse momentum space.}

\FullConference{The XVIth Quark Confinement and the Hadron Spectrum Conference (QCHSC24)\\
 19-24 August, 2024\\
 Cairns Convention Centre, Cairns, Queensland, Australia\\}

\begin{document}
\maketitle

\section{Introduction}

Various theoretical frameworks have been developed to explore different aspects of hadronic partonic structure. Notable approaches include discretized space-time Euclidean lattice calculations~\cite{Hagler:2009ni,Joo:2019byq,MILC:2009mpl,BMW:2008jgk} and the Dyson-Schwinger equations of QCD~\cite{Maris:2003vk,Roberts:1994dr,Bashir:2012fs}. Significant advancements have also been made using the Hamiltonian formulation of QCD quantized on the light front (LF)~\cite{Brodsky:1997de}. Additionally, light-front holography provides complementary insights into nonperturbative QCD~\cite{Brodsky:2014yha}. Basis light-front quantization (BLFQ), a Hamiltonian-based approach, offers a nonperturbative framework for solving relativistic many-body bound-state problems in quantum field theories~\cite{Vary:2009gt,Zhao:2014xaa,Nair:2022evk,Wiecki:2014ola,Li:2015zda,Jia:2018ary,Lan:2019vui,Mondal:2019jdg,Xu:2021wwj,Kuang:2022vdy,Lan:2021wok,Xu:2022yxb,Xu:2024sjt}.

We solve for the mass eigenstates of the LFQCD Hamiltonian within the BLFQ framework~\cite{Vary:2009gt}. The Hamiltonian, formulated with quarks ($q$) and gluons ($g$) as explicit degrees of freedom, incorporates fundamental QCD interactions relevant to the constituent Fock sectors of the nucleon: $|qqq\rangle$, $|qqqg\rangle$, and $|qqqq\bar{q}\rangle$~\cite{Brodsky:1997de}. From the resulting wave functions, obtained as eigenvectors of the Hamiltonian, we compute the nucleon's electromagnetic form factors (FFs), parton distribution functions (PDFs), and axial and tensor charges. The Fourier transform of the FFs reveals spatial distributions of the nucleon’s constituents, such as charge and magnetization distributions. The PDFs describe the nonperturbative structure of the nucleon by encoding the number densities of its confined constituents as functions of the longitudinal momentum fraction ($x$). Our approach, based on a truncated Fock space, is well-suited for low energy scales, and we apply QCD evolution to the PDFs to extend comparisons to higher momentum scales relevant for global analyses. 

We investigate the contributions of quark and gluon helicity, including the contributions arising from sea quarks, to nucleon spin. The RHIC spin program at BNL has established that gluon helicity ($\Delta G$) is nonzero and likely substantial~\cite{STAR:2014wox,deFlorian:2014yva,Nocera:2014gqa,Ethier:2017zbq}, complementing the known quark helicity contribution of approximately 30\%. This underscores the significant role of parton helicities in nucleon spin. However, large uncertainties remain in the small-$x$ contribution to $\Delta G$, defined as the first moment of the polarized gluon PDF~\cite{Ji:2020ena}. Resolving this is a major objective of future Electron-Ion Colliders (EICs)~\cite{Accardi:2012qut,AbdulKhalek:2021gbh,Anderle:2021wcy}. A comprehensive understanding of nucleon distribution functions in both longitudinal and transverse momentum space, along with its spin structure, necessitates a unified framework. Here, we present such a framework, which successfully captures the qualitative features of nucleon properties at the hadronic scale.

\section{Nucleon wave functions from LFQCD Hamiltonian}

The nucleon's structural information is encapsulated in its light-front wave functions (LFWFs), which are derived by solving the Hamiltonian eigenvalue equation:
\begin{align}
P^- P^+ |{\Psi}\rangle = M^2 |{\Psi}\rangle,
\end{align}
where $P^\pm = P^0 \pm P^3$ represent the longitudinal momentum ($P^+$) and the light-front Hamiltonian ($P^-$) of the system, with $M^2$ as the squared mass eigenvalue. At a fixed light-front time, the nucleon state can be expressed as
\begin{align}\label{eq:Fock_space}
|\Psi\rangle=&\psi^{(3q)}|qqq\rangle+\psi^{(3q+g)}|qqqg\rangle + \psi^{(3q+u\bar{u})}|qqqu\bar{u}\rangle\nonumber\\& + \psi^{(3q+d\bar{d})}|qqqd\bar{d}\rangle + \psi^{(3q+s\bar{s})}|qqqs\bar{s}\rangle + \dots\, , 
\end{align}
where $\psi^{(\dots)}$ denotes the probability amplitudes for different parton configurations within the nucleon. These amplitudes can be used to define the LFWFs in either coordinate or momentum space.

The LFQCD Hamiltonian, which includes interactions relevant to the Fock components in Eq.\eqref{eq:Fock_space} formulated in the LF gauge, is given by~\cite{Brodsky:1997de}: 
\begin{align}\label{eqn:PQCD}
P^-_{\rm QCD}=&\int {\rm d}^2x^\perp {\rm d}x^-\Big\{\frac{1}{2}\bar{\Phi}\gamma^+\frac{(m_q+\delta m_q)^2+(i\partial^\perp)^2}{i\partial^+}\Phi+\frac{1}{2}A_a^\mu[\delta m_{g}^2+(i\partial^\perp)^2]A_\mu^a\nonumber\\
&+g_s\bar{\Phi}\gamma^\mu T^aA^a_\mu\Phi+\frac{g_s^2}{2}\bar{\Phi}\gamma^+T^a\Phi\frac{1}{(i\partial^+)^2}\bar{\Phi}\gamma^+T^a\Phi+\frac{g_s^2}{2}\bar{\Phi}\gamma^\mu T^a A_\mu^a\frac{\gamma^+}{i\partial^+}\gamma^\nu T^b A_\nu^b\Phi\Big\}.
\end{align}
Here, $\Phi$ and $A^\mu$ represent the quark and gluon fields, respectively, while $x^-$ and $x^{\perp}$ denote the longitudinal and transverse position coordinates. The generator of the $SU(3)$ gauge group in color space is given by $T$, and $\gamma^\mu$ are the Dirac matrices. In Eq.~\eqref{eqn:PQCD}, the first two terms correspond to the kinetic energies of quarks with physical mass $m_q$ and gluons with zero mass, respectively. The last three terms describe vertex and instantaneous interactions, governed by the global coupling constant $g_s$.
A Fock sector-dependent renormalization procedure, originally developed for positronium in the $|e\bar{e}\rangle$ and $|e\bar{e}\gamma\rangle$ bases~\cite{Zhao:2014hpa,Zhao:2020kuf} and later applied to hadrons~\cite{Lan:2021wok,Xu:2022yxb}, is used to determine the quark and gluon mass counterterms, $\delta m_q$ and $\delta m_g$. Additionally, we introduce a distinct quark mass $m_f$ to incorporate nonperturbative effects in vertex interactions~\cite{Burkardt:1998dd,Glazek:1992aq}.

We employ the BLFQ framework~\cite{Vary:2009gt} to solve the Hamiltonian eigenvalue problem, which has been successfully applied to describe the structure of light mesons~\cite{Lan:2021wok} and nucleons~\cite{Xu:2022yxb} by incorporating dynamical gluons in Fock space. While previous studies~\cite{Lan:2021wok,Xu:2022yxb} introduced an effective confinement mechanism and a substantial gluon mass to mimic confinement effects, our approach here omits the effective confining potential and instead adopts the physical gluon mass.

For the longitudinal direction, we use a plane-wave basis confined within a one-dimensional box of length $2L$, imposing antiperiodic (periodic) boundary conditions for quarks (gluons). In the transverse direction, we employ a two-dimensional harmonic oscillator ("2D-HO") wave function, $\Phi_{nm}(\vec{p}_\perp; b)$, with scale parameter $b$, and incorporate light-cone helicity states in spin space~\cite{Zhao:2014xaa}. This basis formulation transforms the Hamiltonian eigenvalue problem into a matrix eigenvalue problem. Each single-particle parton state is characterized by four quantum numbers: $\bar{\alpha} = \{k, n, m, \lambda\}$. 
Here, $k$ denotes the longitudinal degree of freedom, corresponding to the parton longitudinal momentum $p^+ = \frac{2\pi k}{L}$, where $k$ takes positive half-integer values for quarks and integer values for gluons; the gluon zero mode is omitted. The 2D-HO wave function is characterized by the principal quantum number $n$, the orbital angular momentum quantum number $m$, and the spin $\lambda$. For Fock sectors containing multiple color-singlet states, additional labels are introduced to distinguish them. Specifically, the $|qqqg\rangle $ sector has two color-singlet states, while the $ |qqqq\bar{q}\rangle $ sector contains three. Each many-parton state in a given Fock sector is constructed as a direct product of a specifically ordered flavor-space-spin configuration with a global color-singlet state. Future refinements will ensure total antisymmetry with higher precision.

To ensure a finite matrix representation, we impose two basis space truncations: $ N_{\rm max} $ and $ K $ \cite{Zhao:2014xaa}. The truncation $ N_{\rm max} $ limits the total HO energy of the basis states in the transverse direction as  
\begin{align}
\sum_i \left( 2 n_i + |m_i| + 1 \right) \leq N_{\rm max}.
\end{align} 
The basis cutoff \( N_{\text{max}} \) acts as an implicit regulator for the LFWFs in the transverse direction. The infrared (IR) and ultraviolet (UV) cutoffs are given by
\begin{align}
\Lambda_{\rm IR} \sim \frac{b}{\sqrt{N_{\text{max}}}}
\quad \text{and} \quad
\Lambda_{\rm UV} \sim b \sqrt{N_{\text{max}}},
\end{align} 
respectively. The truncation \( K \) constrains the total longitudinal momentum,  
\begin{align}
\sum_i k_i = K,
\end{align} 
where $ x_i = k_i / K $ represents the longitudinal momentum fraction. The parameter $ K $ thus controls the longitudinal resolution and, consequently, the resolution of the PDFs.

The nucleon LFWFs with helicity $\Lambda$ in momentum space are expressed as a sum over contributions from each Fock sector. These wave functions are given by:

\begin{equation}
\Psi^{\mathcal{N},\,\Lambda}_{\{x_i,\vec{p}_{\perp i},\lambda_i\}} = \sum_{\{n_i m_i\}} \psi^{\mathcal{N}}(\{\overline{\alpha}_i\}) \prod_{i=1}^{\mathcal{N}} \phi_{n_i m_i}(\vec{p}_{\perp i}, b)\,,
\label{eqn:wf}
\end{equation}
where $\psi^{\mathcal{N}}(\{\overline{\alpha}_i\})$ represents the wave function component for a specific Fock sector with particle number $\mathcal{N}$. The Fock sectors considered include $|qqq\rangle$, $|qqqg\rangle$, and $|qqqq\bar{q}\rangle$. The functions $\phi_{n_i m_i}(\vec{p}_{\perp i}, b)$ describe the transverse momentum dependence for each constituent particle.

\section{Results and discussions}
All calculations are performed using $N_{\rm max} = 7$ and $K = 16.5$. The harmonic oscillator scale parameter is chosen as $b = 0.6$ GeV, and the ultraviolet (UV) cutoff for the instantaneous interaction is set to $b_{\rm inst} = 2.80 \pm 0.15$ GeV. The model parameters $\{m_u, m_d, m_f, g_s\} = \{1.0, 0.85, 5.45 \pm 0.4, 2.90 \pm 0.1\}$ (with all masses in GeV while the coupling constant $g_s$ is dimensionless) are determined by fitting the proton mass and its electromagnetic properties. The relatively large constituent quark masses in the first two Fock sectors, $|qqq\rangle$ and $|qqqg\rangle$, partially account for confinement effects and contribute to the nucleon mass as a QCD bound state with significant binding energy. Additionally, a mass difference of $0.15$ GeV between the up and down quarks is incorporated.

In the $|qqqq\bar{q}\rangle$ Fock sector, the partons are treated as free particles due to the omission of higher Fock sectors in the model. Consequently, the current quark masses ($m_u = 0.0022$ GeV, $m_d = 0.0047$ GeV, $m_s = 0.094$ GeV) are used in this sector instead of the model quark masses. At the model scale, the probabilities for finding the proton in each Fock sector are as follows: $53.10\%$ in $|qqq\rangle$, $26.53\%$ in $|qqqg\rangle$, $8.52\%$ in $|qqqu\bar{u}\rangle$, $8.56\%$ in $|qqqd\bar{d}\rangle$, and $3.29\%$ in $|qqqs\bar{s}\rangle$.

\begin{figure}
\begin{center}
\includegraphics[width=0.5\linewidth]{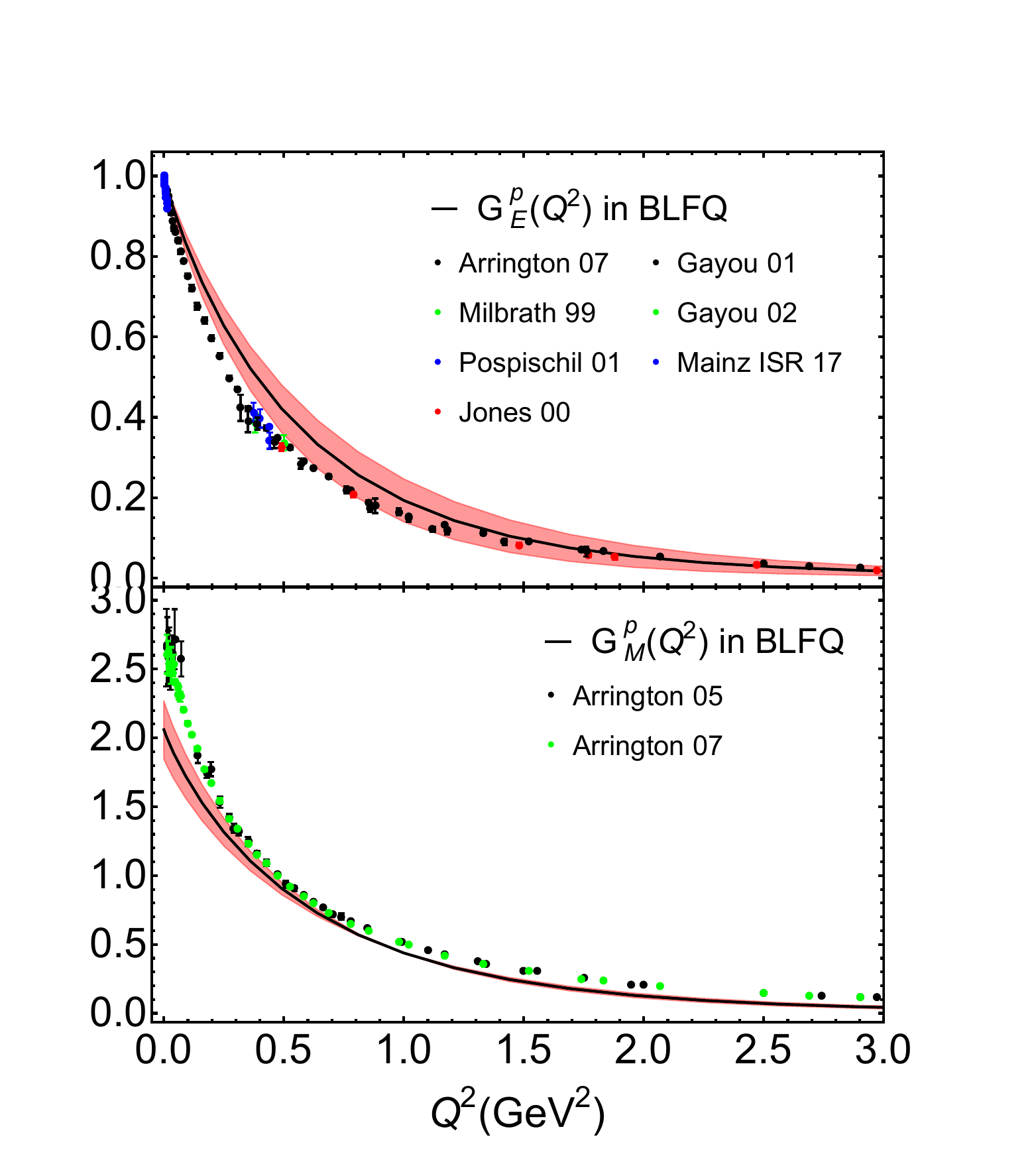}
\caption{Proton electromagnetic form factors. Our results (red bands) are compared with experimental data for $G^{p}_{\rm E}$ from Refs.~\cite{Gayou:2001qt,JeffersonLabHallA:1999epl,Arrington:2007ux,JeffersonLabHallA:2001qqe,A1:2001xxy,BatesFPP:1997rpw}  and for $G^{p}_{\rm M}$ from Refs.~\cite{Arrington:2004ae,Arrington:2007ux}.}
\label{proton_FFs}
\end{center}
\end{figure}
\subsection{Electromagnetic form factors}
Using the obtained LFWFs, the flavor-dependent Dirac $F^q_1(Q^2)$ and Pauli $F^q_2(Q^2)$ FFs of the proton can be expressed as overlap integrals~\cite{Brodsky:2000xy}:
\begin{align}\label{eq_DF}
F_1^q(Q^2) &= \frac{1}{2} \int_{\mathcal{N}} \Psi^{\mathcal{N},\,\Lambda\,*}_{\{x_i^\prime,\vec{p}_{\perp i}^{\,\prime},\lambda_i\}} \, \Psi^{\mathcal{N},\,\Lambda}_{\{x_i,\vec{p}_{\perp i},\lambda_i\}}, \\
F_2^q(Q^2) &= -\frac{M}{(q^1-iq^2)} \int_{\mathcal{N}} \Psi^{\mathcal{N},\,\Lambda\,*}_{\{x_i^\prime,\vec{p}_{\perp i}^{\,\prime},\lambda_i\}} \, \Psi^{\mathcal{N},\,-\Lambda}_{\{x_i,\vec{p}_{\perp i},\lambda_i\}},
\end{align}
where the integration measure $\int_{\mathcal{N}}$ is defined as:
\[
\int_{\mathcal{N}} \equiv \sum_{\mathcal{N},\,\Lambda,\,\lambda_i} \prod_{i=1}^\mathcal{N} \int \left[\frac{{\rm d}x\,{\rm d}^2\vec{p}_\perp}{16\pi^3}\right]_i 16\pi^3 \, \delta\left(1-\sum x_j\right) \, \delta^2\left(\sum \vec{p}_{\perp j}\right).
\]
The calculations are performed in the frame where the momentum transfer is purely transverse, $q = (0, 0, \vec{q}_{\perp})$, such that $Q^2 = -q^2 = \vec{q}_{\perp}^{\,2}$. For the struck quark, the momentum fractions and transverse momenta are modified as follows: ${x^\prime}_1 = x_1$ and ${\vec{p}}_{\perp 1}^{\,\prime} = \vec{p}_{\perp 1} + (1 - x_1) \vec{q}_\perp$ for the struck quark, while for the spectator partons, ${x^\prime}_i = {x_i}$ and ${\vec{p}}_{\perp i}^{\,\prime} = \vec{p}_{\perp i} - {x_i} \vec{q}_\perp$.

The nucleon form factors are derived from the flavor form factors~\cite{Cates:2011pz}. The proton's Sachs form factors, $G_{\rm E}(Q^2)$ and $G_{\rm M}(Q^2)$, are given by:
\begin{align}
G_{\rm E}(Q^2) &= F_1(Q^2) - \frac{Q^2}{4M^2} F_2(Q^2), \\
G_{\rm M}(Q^2) &= F_1(Q^2) + F_2(Q^2).
\end{align}
These expressions relate the Dirac and Pauli form factors to the experimentally measurable electric and magnetic form factors of the proton.

The electromagnetic FFs of the proton are presented in Fig.~\ref{proton_FFs}. The red bands indicate the uncertainty in our approach, arising from the uncertainties in the model parameters mentioned earlier. Our results demonstrate reasonable agreement with experimental data for the proton electric form factor. For the proton magnetic form factor, the agreement with data is also satisfactory at high $Q^2$, although a deviation of approximately 25$\%$ is observed at low $Q^2$. 

The electromagnetic radii are determined from the slopes of the Sachs form factors~\cite{Ernst:1960zza}. We obtain a charge radius of $\sqrt{\langle r^2_{E}\rangle} = 0.72 \pm 0.05$ fm and a magnetic radius of $\sqrt{\langle r^2_{M}\rangle} = 0.73 \pm 0.02$ fm. These values can be compared to the experimental results: $\sqrt{\langle r^2_{E}\rangle}_{\rm exp} = 0.840^{+0.003}_{-0.002}$ fm and $\sqrt{\langle r^2_{M}\rangle}_{\rm exp} = 0.849^{+0.003}_{-0.003}$ fm~\cite{Lin:2021xrc,ParticleDataGroup:2024cfk}.

\subsection{Parton distribution functions}
Using the LFWFs, the unpolarized, helicity, and transversity PDFs of the nucleon are expressed as follows:
\begin{align}
f(x) &= \int_{\mathcal{N}} \frac{1}{2} \, \Psi^{\mathcal{N},\,\Lambda\,*}_{\{x_i,\vec{p}_{\perp i},\lambda_i\}} \, \Psi^{\mathcal{N},\,\Lambda}_{\{x_i,\vec{p}_{\perp i},\lambda_i\}} \, \delta(x - x_i), \\
\Delta f(x) &= \int_{\mathcal{N}} \frac{\lambda_1}{2} \, \Psi^{\mathcal{N},\,\Lambda\,*}_{\{x_i,\vec{p}_{\perp i},\lambda_i\}} \, \Psi^{\mathcal{N},\,\Lambda}_{\{x_i,\vec{p}_{\perp i},\lambda_i\}} \, \delta(x - x_i), \\
\delta f(x) &= \int_{\mathcal{N}} \Psi^{\mathcal{N},\,\Lambda\,*}_{\{x_i,\vec{p}_{\perp i},\lambda_i^\prime\}} \, \Psi^{\mathcal{N},\,-\Lambda}_{\{x_i,\vec{p}_{\perp i},\lambda_i\}} \, \delta(x - x_i),
\label{eqn:pdf_i}
\end{align}
respectively. For quark distributions, contributions come from both the $\mathcal{N}=3,\,4,\,5$ Fock sectors, whereas for gluon distributions, only the $\mathcal{N}=4$ Fock sector contributes. In Eq.~\eqref{eqn:pdf_i}, $\lambda^\prime_1 = -\lambda_1$ for the struck parton, while $\lambda^\prime_i = \lambda_i$ for the spectator partons ($i \ne 1$). 
The PDFs, interpreted as particle number densities, satisfy the normalization condition for valence quarks: $\int_{0}^1 f(x) \, {\rm d}x = n_q$, where $n_q = 1$ for down quarks and $n_q = 2$ for up quarks. When including gluon and sea quark PDFs, the second moment of the PDFs adheres to the momentum sum rule: $\int_0^1 \sum_i x f^i(x) \, {\rm d}x = 1$.

\begin{figure}[htp]
\begin{center}
\includegraphics[width=0.5\linewidth]{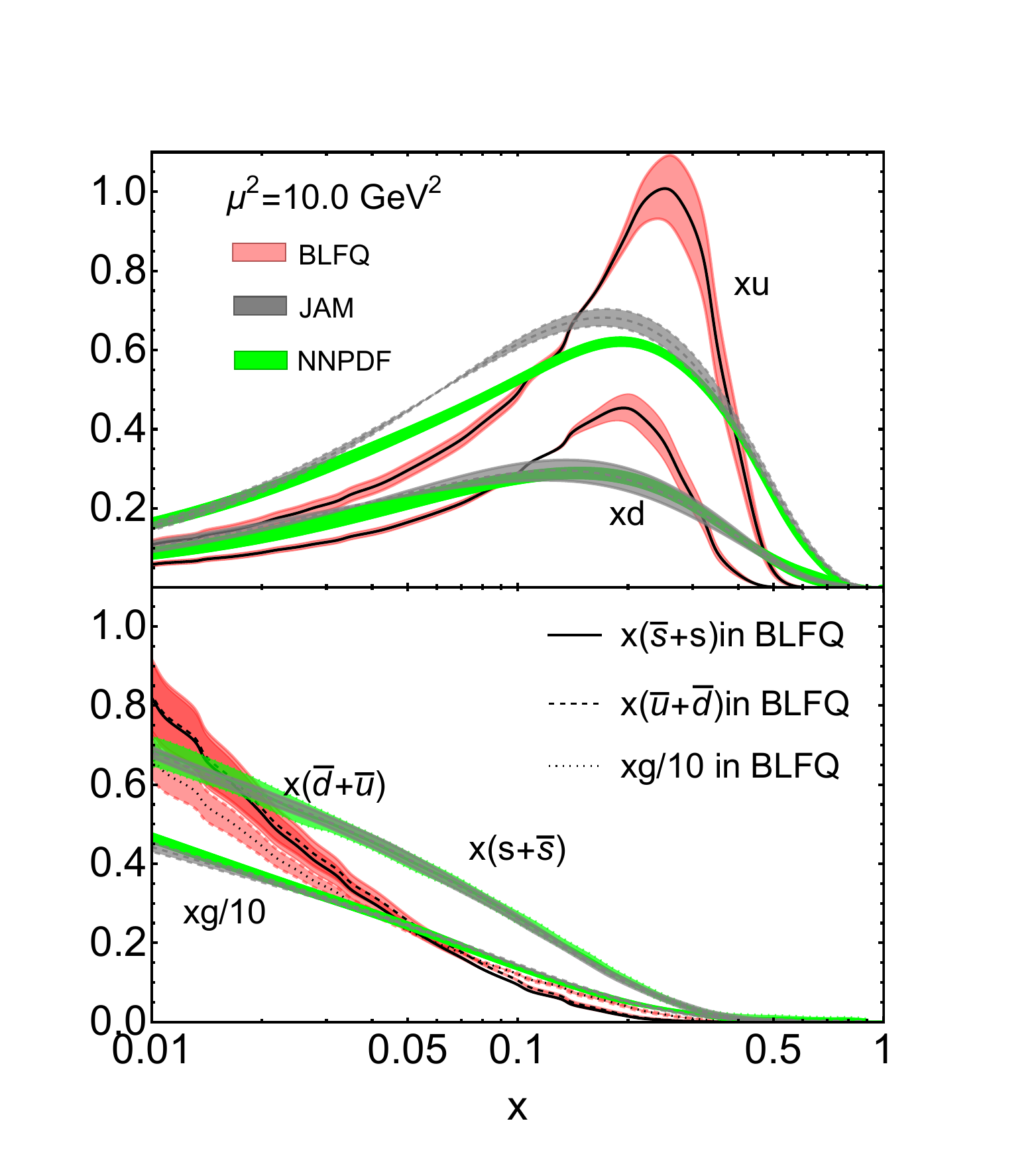}
\caption{Proton unpolarized PDFs. Our results (red bands) are compared with the global analyses from NNPDF3.1~\cite{NNPDF:2017mvq} (light-green bands) and JAM~\cite{Cocuzza:2021cbi} (gray bands). }
\label{fig_pdf}
\end{center}
\end{figure}
To evolve our PDFs from the model scale to a higher scale, we utilize the Higher Order Perturbative Parton Evolution toolkit~\cite{Salam:2008qg}, which numerically solves the Dokshitzer-Gribov-Lipatov-Altarelli-Parisi (DGLAP) equations of QCD~\cite{Dokshitzer:1977sg,Gribov:1972ri,Altarelli:1977zs}. The model scale is determined to be $\mu_0^2 = 0.22 \pm 0.02$ GeV$^2$ by matching the second moment of the total valence quark PDFs at $\mu^2 = 10$ GeV$^2$ with global fitting data, yielding average values of $\langle x\rangle_{u+d} = 0.37 \pm 0.01$~\cite{NNPDF:2017mvq,Cocuzza:2021cbi}.

Figure~\ref{fig_pdf} presents our results for the proton's unpolarized PDFs at $\mu^2 = 10$ GeV$^2$, comparing the valence quark and gluon distributions after QCD evolution with the NNPDF3.1~\cite{NNPDF:2017mvq} and JAM global analyses~\cite{Cocuzza:2021cbi}. We find qualitative agreement between our quark distributions and the global fits. However, the large constituent quark masses pose challenges in modeling longitudinal excitations, particularly due to the absence of a longitudinal confining potential, which plays a significant role in these excitations. As a result, our valence quark PDFs are narrower than the global fits in the region $x > 0.1$. 

The gluon PDF shows good consistency with the global fits in the region $x > 0.05$. In the small-$x$ region, where QCD evolution is highly sensitive to the model scale $\mu_0$, our gluon PDF qualitatively aligns with the global analyses.

\begin{figure}
\begin{center}
\includegraphics[width=0.6\linewidth]{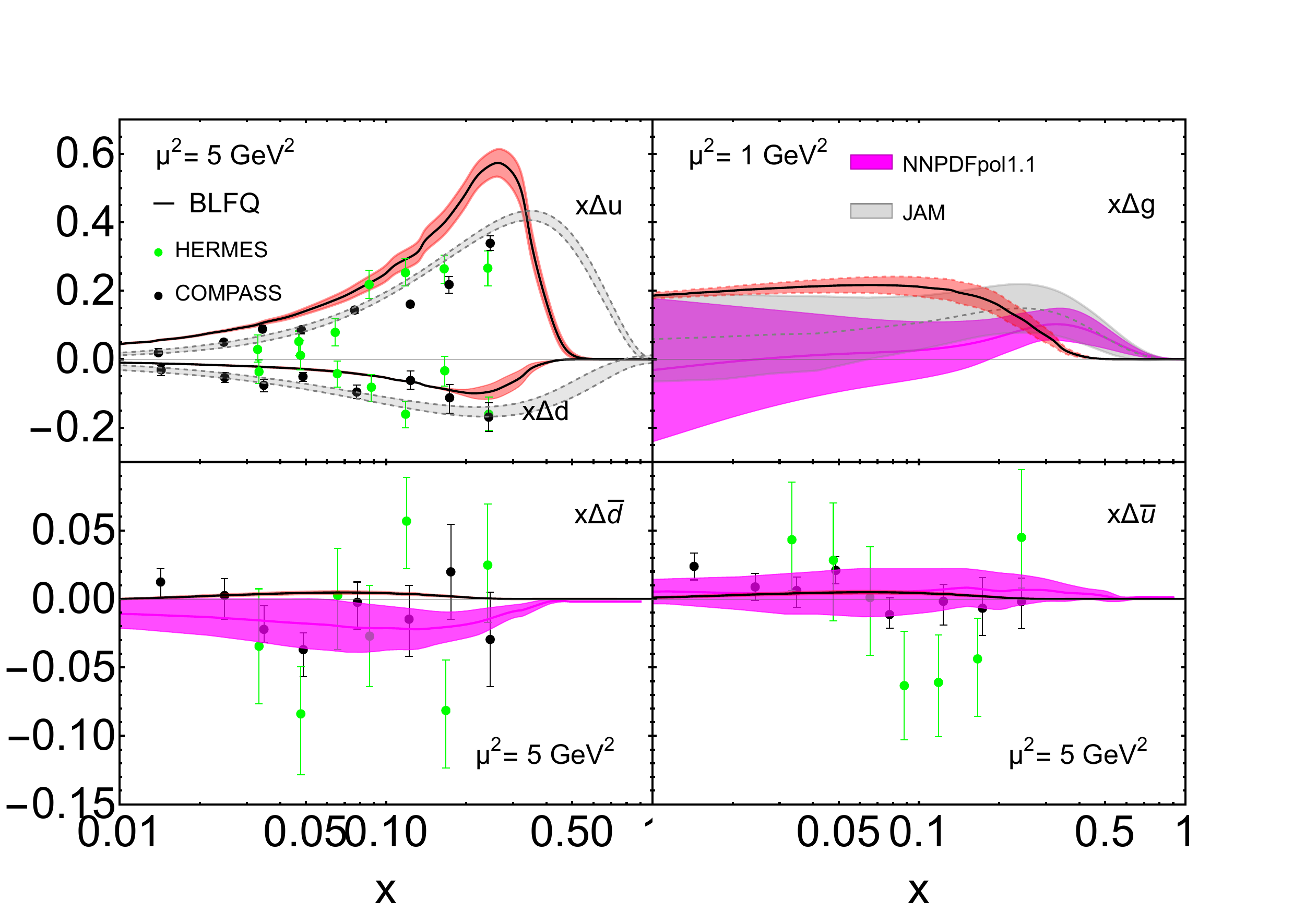}
\caption{The helicity PDFs of the valence $u$ and $d$ (upper-left), $g$ (upper-right), $\bar{u}$ (lower-left), and $\bar{d}$ (lower-right) in the proton. The experimental data are taken from COMPASS~\cite{COMPASS:2010hwr} and HERMES~\cite{HERMES:2003gbu,HERMES:2004zsh} Collaborations. The magenta and gray bands represent the global analysis by NNPDFpol1.1~\cite{Nocera:2014gqa} and JAM~\cite{Sato:2016tuz}, respectively.}
\label{fig_hpdf}
\end{center}
\end{figure}

Figure~\ref{fig_hpdf} shows the helicity PDFs for valence up and down quarks (upper-left), gluons (upper-right), up-sea quarks (lower-left), and down-sea quarks (lower-right). Our results for the quark helicity PDFs exhibit rough agreement with experimental data from COMPASS~\cite{COMPASS:2010hwr} and HERMES~\cite{HERMES:2003gbu,HERMES:2004zsh}. For comparison, we also include the global analyses from NNPDFpol1.1~\cite{Nocera:2014gqa} and JAM~\cite{Sato:2016tuz}. Similar to the unpolarized PDFs, our valence quark helicity distributions are narrower than the global fits in the region $x > 0.1$. It is worth noting that the signs of the sea quark helicity PDFs remain undetermined by experimental data.

\begin{figure}
\begin{center}
\includegraphics[width=0.5\linewidth]{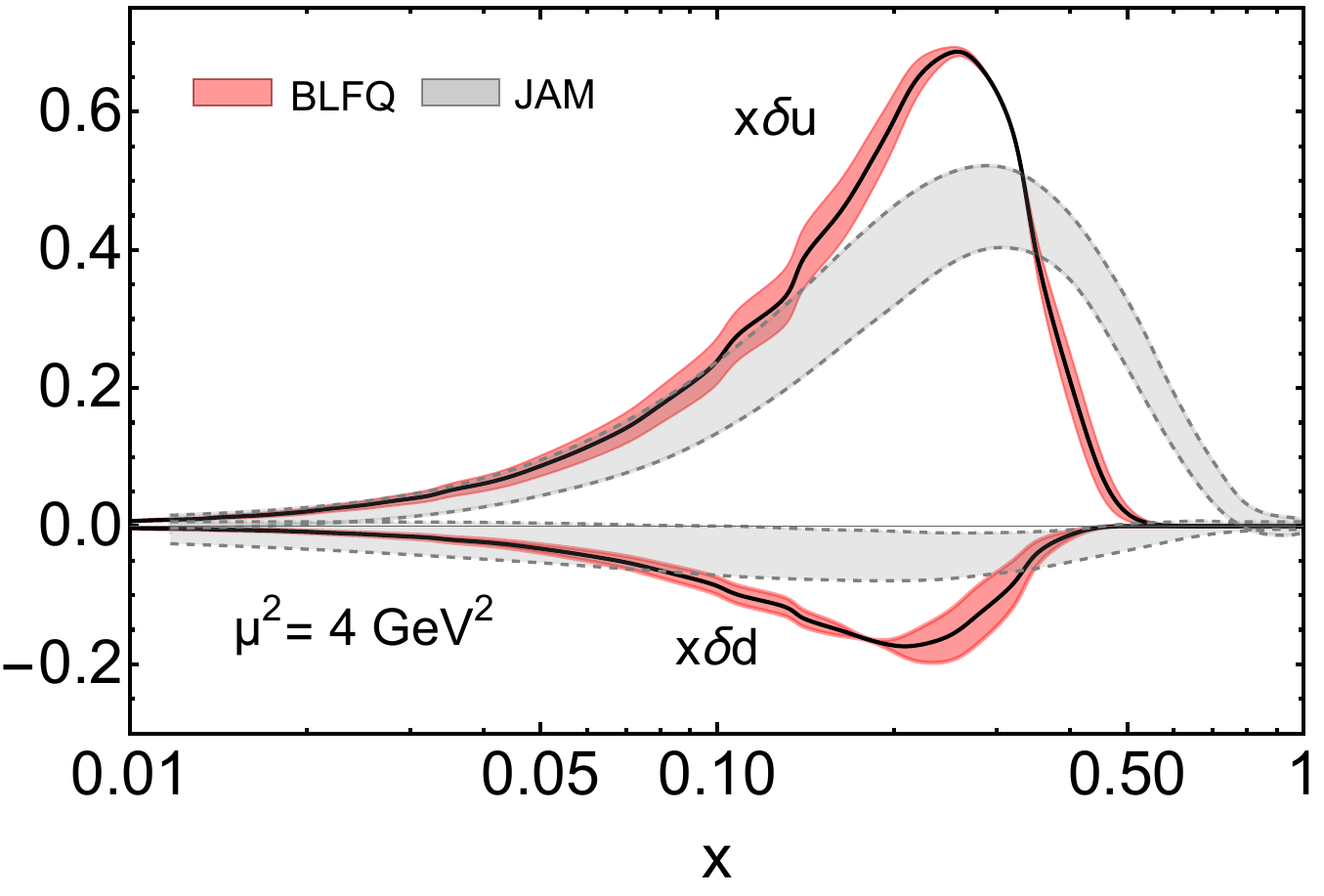}
\caption{Comparison for the transversity PDF in the proton from BLFQ (red
bands) and the recent global analyses by JAM~\cite{Cocuzza:2023oam} (gray band).
}
\label{fig_h1}
\end{center}
\end{figure}

We also present the gluon helicity PDF in Fig.~\ref{fig_hpdf} at the scale $\mu^2 = 1$ GeV$^2$ and compare our predictions with the global analyses by JAM~\cite{Sato:2016tuz} and the NNPDF Collaboration~\cite{Nocera:2014gqa}. Our results show rough agreement with the global fits in the small-$x$ region, while our gluon helicity distribution decreases more rapidly at large-$x$ compared to the global analyses. Our predictions align more closely with the JAM results up to $x \sim 0.2$. It is important to note that significant uncertainties remain in the global analyses, particularly in the small-$x$ region, where even the sign of the gluon helicity distribution is uncertain~\cite{Zhou:2022wzm}.

The partonic helicity contributions to the proton spin are given by the first moment of the helicity distributions, known as the axial charge. At the model scale, our analysis yields axial charges of $\Delta \Sigma_u = 0.89 \pm 0.07$ for the up quark and $\Delta \Sigma_d = -0.22 \pm 0.02$ for the down quark. Our prediction for $\Delta \Sigma_u$ is in good agreement with the world data summarized in Ref.~\cite{Deur:2018roz}, while $\Delta \Sigma_d$ is slightly smaller in magnitude than most global extractions. Consequently, the total quark helicity contribution to the proton spin, $\frac{1}{2} \Delta \Sigma = 0.33 \pm 0.04$, is somewhat larger than the world data.  

Additionally, we find a significant gluon contribution to the proton spin, $\Delta G = 0.29 \pm 0.03$ for $x_g \in [0.05, 0.2]$ at $10$ GeV$^2$, which is consistent with the NNPDF analysis $\Delta G = 0.23(6)$~\cite{Nocera:2014gqa} and the lattice QCD prediction at the physical pion mass $\Delta G = 0.251(47)(16)$~\cite{Yang:2016plb}.  

In the light-cone gauge, the partonic orbital angular momentum (OAM) contributions to the proton spin can be computed using the generalized parton distribution (GPD) sum rule~\cite{Ji:1996ek}:  
\begin{equation}
L_z = \int {\rm d}x \left\{ \frac{1}{2} x [H(x,0,0) + E(x,0,0)] - \tilde{H}(x,0,0) \right\}.
\end{equation}
Our numerical results give $L_z^u = 0.036 \pm 0.002$, $L_z^d = 0.058 \pm 0.001$, and $L_z^g = -0.037 \pm 0.004$.  
Currently, no direct experimental measurement of OAM exists. However, gluon OAM can be accessed through double spin asymmetry in diffractive dijet production~\cite{Bhattacharya:2022vvo}, while quark OAM can be probed in exclusive double Drell-Yan processes~\cite{Bhattacharya:2017bvs}.

The transversity PDF quantifies the degree of transverse polarization of quarks within a transversely polarized nucleon. In Fig.~\ref{fig_h1}, we compare our predictions for the transversity PDFs with the recent JAM analysis at $\mu^2 = 4$ GeV$^2$~\cite{Cocuzza:2023oam}. Our results agree with the global fit in the region $x < 0.1$. However, we observe a narrow peak at large $x$ for both up and down quarks, a feature also seen in the unpolarized and helicity PDFs. Additionally, the transversity and helicity PDFs exhibit a symmetric behavior for up and down quarks, a symmetry that was disrupted by the effective confining potential in our earlier study~\cite{Mondal:2019jdg}. We propose that this symmetry could be influenced by incorporating higher Fock sectors and additional QCD interactions. It is important to note that the transversity PDFs for gluons cannot be defined in the same way as for quarks.

The tensor charge is derived from the first moment of the transversity PDFs. At the scale $\mu^2 = 4$ GeV$^2$, our calculations yield a tensor charge of $\delta u = 0.81 \pm 0.08$ for the up quark, which is slightly larger than the recent JAM analysis result of $\delta u = 0.71 (2)$~\cite{Cocuzza:2023oam} and the lattice QCD prediction of $\delta u = 0.784 (28)$~\cite{Gupta:2018lvp}. For the down quark, our prediction of $\delta d = -0.22 \pm 0.01$ is in good agreement with the recently extracted value of $\delta d = -0.200 (6)$~\cite{Cocuzza:2023oam} and the lattice QCD result of $\delta d = -0.204 (11)$~\cite{Gupta:2018lvp}. It is worth noting that these observables can, in principle, be systematically improved by including higher Fock components that will naturally incorporate additional QCD vertices in the Hamiltonian.

\begin{figure}
\begin{center}
\includegraphics[width=0.3\linewidth]{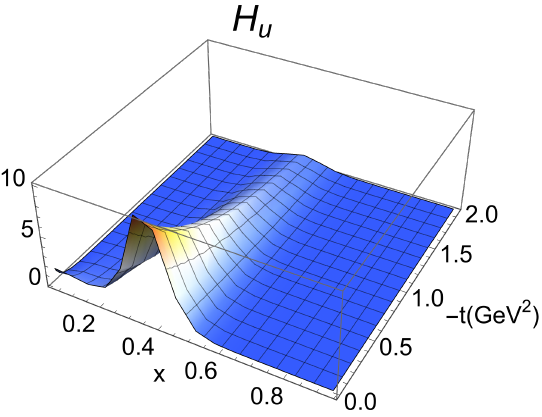}
\includegraphics[width=0.3\linewidth]{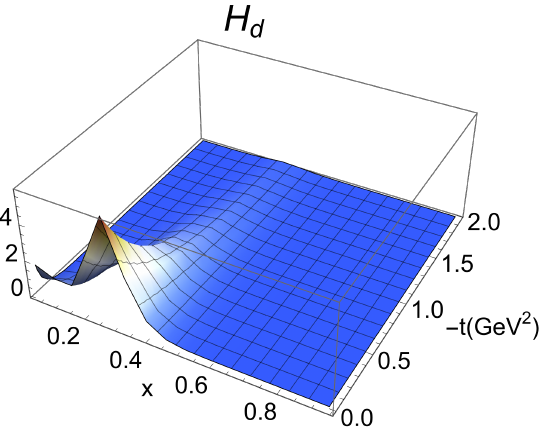}
\includegraphics[width=0.31\linewidth]{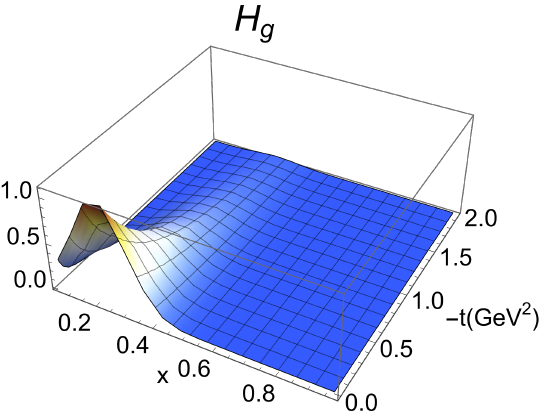}\\
\includegraphics[width=0.3\linewidth]{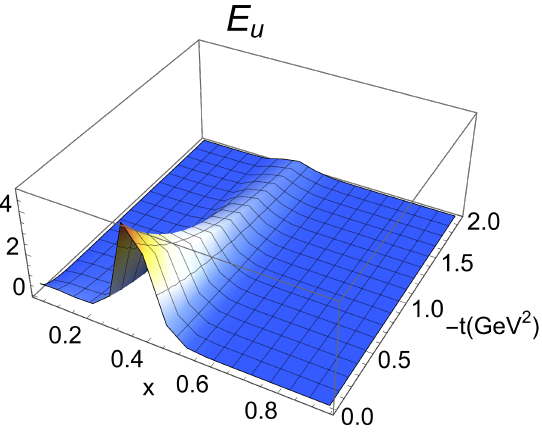}
\includegraphics[width=0.3\linewidth]{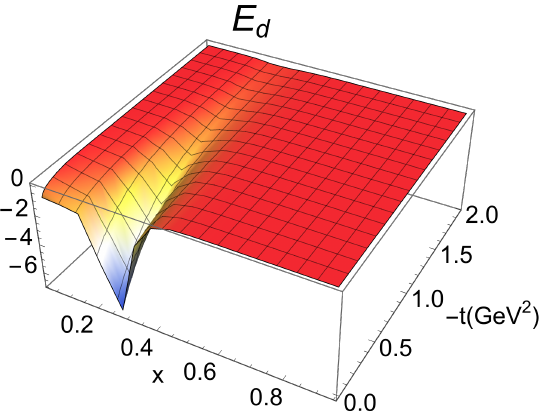}
\includegraphics[width=0.31\linewidth]{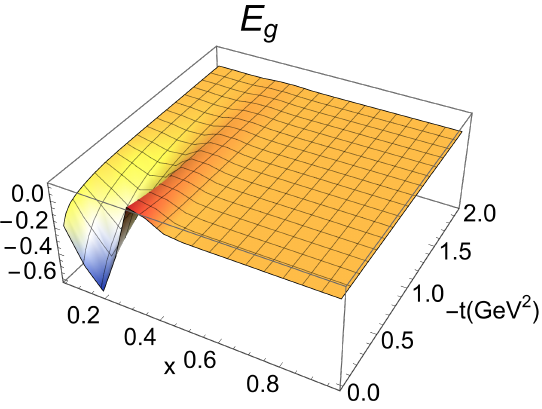}
\caption{The unpolarized GPDs: $H(x,0,t)$ and $E(x,0,t)$ for the $u$ quark (left panel), $d$ quark (middle panel), and gluon (right pannel). The GPDs are presented with respect to $x$ and $-t$.}
\label{GPDs}
\end{center}
\end{figure}
\subsection{Generalized parton distribution functions}
Multivariable GPDs~\cite{Diehl:2003ny} depend on $(x, \xi, t)$, where $x$ is the longitudinal momentum fraction carried by the parton, $\xi$ (defined as $-q^+/2P^+$) represents the longitudinal momentum transfer, and $t$ (defined as $q^2$) is the square of the total momentum transferred. Although GPDs are not probabilistic, their two-dimensional Fourier transforms from transverse momentum transfer to the impact-parameter plane—in the absence of longitudinal momentum transfer—provide a probabilistic interpretation of GPDs~\cite{Burkardt:2002hr}.
The overlap representation of the unpolarized GPDs in terms of the LFWFs for $\xi=0$  is expressed as
\begin{align}\label{eq_DF}
H^{q/g}(x,0,t) &= \frac{1}{2} \int_{\mathcal{N}} \Psi^{\mathcal{N},\,\Lambda\,*}_{\{x_i^\prime,\vec{p}_{\perp i}^{\,\prime},\lambda_i\}} \, \Psi^{\mathcal{N},\,\Lambda}_{\{x_i,\vec{p}_{\perp i},\lambda_i\}} \delta (x-x_1), \\
E^{q/g}(x,0,t) &= -\frac{M}{(q^1-iq^2)} \int_{\mathcal{N}} \Psi^{\mathcal{N},\,\Lambda\,*}_{\{x_i^\prime,\vec{p}_{\perp i}^{\,\prime},\lambda_i\}} \, \Psi^{\mathcal{N},\,-\Lambda}_{\{x_i,\vec{p}_{\perp i},\lambda_i\}} \delta (x-x_1).
\end{align}

We present our  unpolarized GPDs $H$ and $E$ for each quark flavor and for the gluon at the model scale in Fig.~\ref{GPDs}, plotting them against the light-cone momentum fraction $x$ and the squared momentum transfer $-t$. The distributions peak when the proton transfers no transverse momentum and the struck quark carries less than 50\% of the proton’s longitudinal momentum. As transverse momentum transfer increases, the peak shifts to higher $x$ values while the magnitude decreases. At large $x$, all distributions decay and become $t$-independent, with $E$ decaying faster than $H$. Due to the negative anomalous magnetic moment of the $d$ quark, $E^d$ is negative. These features, consistent across QCD-inspired models~\cite{Mondal:2015uha,Pasquini:2006dv,Chakrabarti:2013gra,deTeramond:2018ecg,Lin:2023ezw,Kaur:2023lun,Liu:2024umn} and lattice QCD~\cite{Alexandrou:2020zbe,Lin:2020rxa,Bhattacharya:2022aob}, appear model-independent. However, the large constituent quark masses make modeling longitudinal excitations challenging, primarily due to the absence of higher Fock sectors or a phenomenological longitudinal confining potential, either of which could strongly influence for such excitations. Consequently, our distributions are narrower than those from QCD-inspired models and lattice QCD for $x > 0.1$.
\section{Conclusion}

We have solved the LFQCD Hamiltonian for the nucleon within a combined Fock space framework, including the three-quark ($|qqq\rangle$), three-quark-one-gluon ($|qqqg\rangle$), and three-quark-one-quark-antiquark ($|qqqq\bar{q}\rangle$) sectors, using the BLFQ approach. This has enabled us to obtain LFWFs for computing the nucleon's partonic structure. 

We have calculated the electromagnetic FFs and the unpolarized, helicity, and transversity PDFs for quarks and gluons and the unpolarized GPDs. Our results show qualitative agreement with experimental data and global fits. Additionally, we have evaluated the quark and gluon helicity and OAM contributions to the proton spin sum rule. Our predictions indicate that quark helicity contributes $66\%$ and gluon helicity $12\%$ to the proton's total angular momentum. The OAM contributions are found to be $L_z^u = 0.036 \pm 0.002$, $L_z^d = 0.058 \pm 0.001$, and $L_z^g = -0.037 \pm 0.004$. Significant uncertainties remain in the gluon helicity distribution, particularly in the small-$x$ region, highlighting the importance of future measurements in the domain $x_g < 0.02$. Currently, there is no experimental data on OAM, and the needed experiments constitute a key objective for future EICs. 

Our approach yields tensor charges of $\delta u = 0.81 \pm 0.08$ and $\delta d = -0.22 \pm 0.01$. Our predicted values align well with the recent global analysis and lattice QCD predictions. Although our current results for parton distributions, particularly at large-$x$, are less satisfactory, they can be systematically improved. Notably, the inclusion of additional Fock sectors with multi-gluon configurations, which are relevant for three-gluon and four-gluon interactions, is expected to play a crucial role in generating color confinement and improving the agreement between theory and data.

The obtained LFWFs represent a significant step forward in achieving a unified description of various hadron distribution functions in both longitudinal and transverse momentum space. These wave functions can be further utilized to compute quark and gluon TMDs, Wigner distributions, and double parton correlations within the nucleon.

\section*{Acknowledgement} 
C. M. is also supported by new faculty start up funding by the Institute of Modern Physics, Chinese Academy of Sciences, Grant No. E129952YR0. J. Lan is supported by Special Research Assistant Funding Project, Chinese Academy of Sciences, by the National Natural Science Foundation of China (NSFC) under Grant No. 12305095, and the Natural Science Foundation of Gansu Province, China, Grant No. 23JRRA631. X. Zhao is supported by new faculty startup funding by the Institute of Modern Physics, Chinese Academy of Sciences, by Key Research Program of Frontier Sciences, Chinese Academy of Sciences, Grant No. ZDBS-LY-7020, by the Foundation for Key Talents of Gansu Province, by the Central Funds Guiding the Local Science and Technology Development of Gansu Province, Grant No. 22ZY1QA006, by international partnership program of the Chinese Academy of Sciences, Grant No. 016GJHZ2022103FN, by National Key R\&D Program of China, Grant No. 2023YFA1606903, by the Strategic Priority Research Program of the Chinese Academy of Sciences, Grant No. XDB34000000, and by the NSFC under Grant No.12375143. YL is supported by the new faculty startup fund of University of Science and Technology of
China, by the NSFC under Grant No. 12375081, by the Chinese Academy of Sciences under Grant
No. YSBR-101. J. P. V. is supported by the Department of Energy under Grant No. DE-SC0023692. This work is also supported by Gansu International Collaboration and Talents Recruitment Base of Particle Physics (2023–2027), by the Senior Scientist Program funded by Gansu Province Grant No. 25RCKA008.



\begin{thebibliography}{99}

\bibitem{Hagler:2009ni}
P.~Hagler,
Phys. Rept. \textbf{490}, 49-175 (2010).
\bibitem{Joo:2019byq}
B.~Jo\'o \textit{et al.} [USQCD],
Eur. Phys. J. A \textbf{55}, no.11, 199 (2019).

\bibitem{MILC:2009mpl}
A.~Bazavov \textit{et al.} [MILC],
Rev. Mod. Phys. \textbf{82}, 1349-1417 (2010).

\bibitem{BMW:2008jgk}
S.~Durr \textit{et al.} [BMW],
Science \textbf{322}, 1224-1227 (2008).

\bibitem{Maris:2003vk}
P.~Maris and C.~D.~Roberts,
Int. J. Mod. Phys. E \textbf{12}, 297-365 (2003).

\bibitem{Roberts:1994dr}
C.~D.~Roberts and A.~G.~Williams,
Prog. Part. Nucl. Phys. \textbf{33}, 477-575 (1994).

\bibitem{Bashir:2012fs}
A.~Bashir, L.~Chang, I.~C.~Cloet {\it et al.},
Commun. Theor. Phys. \textbf{58}, 79-134 (2012).

\bibitem{Brodsky:1997de}
S.~J.~Brodsky, H.~C.~Pauli and S.~S.~Pinsky,
Phys. Rept. \textbf{301}, 299-486 (1998).

\bibitem{Brodsky:2014yha}
S.~J.~Brodsky, G.~F.~de Teramond, H.~G.~Dosch and J.~Erlich,
Phys. Rept. \textbf{584}, 1-105 (2015).

\bibitem{Vary:2009gt}
J.~P.~Vary, H.~Honkanen, J.~Li, P.~Maris, S.~J.~Brodsky {\it et al.},
Phys. Rev. C \textbf{81}, 035205 (2010).

\bibitem{Zhao:2014xaa}
X.~Zhao, H.~Honkanen, P.~Maris, J.~P.~Vary and S.~J.~Brodsky,
Phys. Lett. B \textbf{737}, 65-69 (2014).

\bibitem{Nair:2022evk}
S.~Nair \textit{et al.} [BLFQ],
Phys. Lett. B \textbf{827}, 137005 (2022).

\bibitem{Wiecki:2014ola}
P.~Wiecki, Y.~Li, X.~Zhao, P.~Maris and J.~P.~Vary,
Phys. Rev. D \textbf{91}, no.10, 105009 (2015).

\bibitem{Li:2015zda}
Y.~Li, P.~Maris, X.~Zhao and J.~P.~Vary,
Phys. Lett. B \textbf{758}, 118-124 (2016).

\bibitem{Jia:2018ary}
S.~Jia and J.~P.~Vary,
Phys. Rev. C \textbf{99}, no.3, 035206 (2019).

\bibitem{Lan:2019vui}
J.~Lan, C.~Mondal, S.~Jia, X.~Zhao and J.~P.~Vary,
Phys. Rev. Lett. \textbf{122}, no.17, 172001 (2019).

\bibitem{Mondal:2019jdg}
C.~Mondal, S.~Xu, J.~Lan, X.~Zhao, Y.~Li {\it et al.},
Phys. Rev. D \textbf{102}, no.1, 016008 (2020).

\bibitem{Xu:2021wwj}
S.~Xu \textit{et al.} [BLFQ],
Phys. Rev. D \textbf{104}, no.9, 094036 (2021).


\bibitem{Kuang:2022vdy}
Z.~Kuang \textit{et al.} [BLFQ],
Phys. Rev. D \textbf{105}, no.9, 094028 (2022).

\bibitem{Lan:2021wok}
J.~Lan \textit{et al.} [BLFQ],
Phys. Lett. B \textbf{825}, 136890 (2022).

\bibitem{Xu:2022yxb}
S.~Xu \textit{et al.} [BLFQ],
Phys. Rev. D \textbf{108}, no.9, 094002 (2023).

\bibitem{Xu:2024sjt}
S.~Xu, Y.~Liu, C.~Mondal, J.~Lan, X.~Zhao, Y.~Li and J.~P.~Vary,
[arXiv:2408.11298 [hep-ph]].

\bibitem{STAR:2014wox}
L.~Adamczyk \textit{et al.} [STAR],
Phys. Rev. Lett. \textbf{115}, no.9, 092002 (2015).

\bibitem{deFlorian:2014yva}
D.~de Florian, R.~Sassot, M.~Stratmann and W.~Vogelsang,
Phys. Rev. Lett. \textbf{113}, 012001 (2014).

\bibitem{Nocera:2014gqa}
E.~R.~Nocera \textit{et al.} [NNPDF],
Nucl. Phys. B \textbf{887}, 276-308 (2014).

\bibitem{Ethier:2017zbq}
J.~J.~Ethier, N.~Sato and W.~Melnitchouk,
Phys. Rev. Lett. \textbf{119}, no.13, 132001 (2017).

\bibitem{Ji:2020ena}
X.~Ji, F.~Yuan and Y.~Zhao,
Nature Rev. Phys. \textbf{3}, no.1, 27-38 (2021).

\bibitem{Accardi:2012qut}
A.~Accardi, J.~L.~Albacete, M.~Anselmino \textit{et al.}
Eur. Phys. J. A \textbf{52}, no.9, 268 (2016).

\bibitem{AbdulKhalek:2021gbh}
R.~Abdul Khalek, A.~Accardi, J.~Adam, D.~Adamiak \textit{et al.}
Nucl. Phys. A \textbf{1026}, 122447 (2022).

\bibitem{Anderle:2021wcy}
D.~P.~Anderle, V.~Bertone, X.~Cao \textit{et al.}
Front. Phys. (Beijing) \textbf{16}, no.6, 64701 (2021).

\bibitem{Zhao:2014hpa}
X.~Zhao,
Few Body Syst. \textbf{56}, no.6-9, 257-265 (2015).

\bibitem{Zhao:2020kuf}
X.~Zhao, K.~Fu, H.~Zhao and J.~P.~Vary,
PoS \textbf{LC2019}, 090 (2020).

\bibitem{Burkardt:1998dd}
M.~Burkardt,
Phys. Rev. D \textbf{58}, 096015 (1998).

\bibitem{Glazek:1992aq}
S.~D.~Glazek and R.~J.~Perry,
Phys. Rev. D \textbf{45}, 3740-3754 (1992).

\bibitem{Brodsky:2000xy}
S.~J.~Brodsky, M.~Diehl and D.~S.~Hwang,
Nucl. Phys. B \textbf{596}, 99-124 (2001).

\bibitem{Cates:2011pz}
G.~D.~Cates, C.~W.~de Jager, S.~Riordan {\it et al.},
Phys. Rev. Lett. \textbf{106}, 252003 (2011).

\bibitem{Gayou:2001qt}
O.~Gayou, K.~Wijesooriya, A.~Afanasev, M.~Amarian  \textit{et al.}
Phys. Rev. C \textbf{64}, 038202 (2001).

\bibitem{JeffersonLabHallA:1999epl}
M.~K.~Jones \textit{et al.} [Jefferson Lab Hall A],
Phys. Rev. Lett. \textbf{84}, 1398-1402 (2000).

\bibitem{Arrington:2007ux}
J.~Arrington, W.~Melnitchouk and J.~A.~Tjon,
Phys. Rev. C \textbf{76}, 035205 (2007).

\bibitem{JeffersonLabHallA:2001qqe}
O.~Gayou \textit{et al.} [Jefferson Lab Hall A],
Phys. Rev. Lett. \textbf{88}, 092301 (2002).
d

\bibitem{A1:2001xxy}
T.~Pospischil \textit{et al.} [A1],
Eur. Phys. J. A \textbf{12}, 125-127 (2001).

\bibitem{BatesFPP:1997rpw}
B.~D.~Milbrath \textit{et al.},
Phys. Rev. Lett. \textbf{80}, 452 (1998).

\bibitem{Arrington:2004ae}
J.~Arrington,
Phys. Rev. C \textbf{71}, 015202 (2005).

\bibitem{Ernst:1960zza}
F.~J.~Ernst, R.~G.~Sachs and K.~C.~Wali,
Phys. Rev. \textbf{119}, 1105-1114 (1960).

\bibitem{Lin:2021xrc}
Y.~H.~Lin, H.~W.~Hammer and U.~G.~Mei\ss{}ner,
Phys. Rev. Lett. \textbf{128}, no.5, 052002 (2022).

\bibitem{ParticleDataGroup:2024cfk}
S.~Navas \textit{et al.} [Particle Data Group],
Phys. Rev. D \textbf{110}, no.3, 030001 (2024).

\bibitem{NNPDF:2017mvq}
R.~D.~Ball \textit{et al.} [NNPDF],
Eur. Phys. J. C \textbf{77}, no.10, 663 (2017).

\bibitem{Cocuzza:2021cbi}
C.~Cocuzza \textit{et al.} [JAM],
Phys. Rev. D \textbf{104}, no.7, 074031 (2021).

\bibitem{Salam:2008qg}
G.~P.~Salam and J.~Rojo,
Comput. Phys. Commun. \textbf{180}, 120-156 (2009).

\bibitem{Dokshitzer:1977sg}
Y.~L.~Dokshitzer,
Sov. Phys. JETP \textbf{46}, 641-653 (1977).

\bibitem{Gribov:1972ri}
V.~N.~Gribov and L.~N.~Lipatov,
Sov. J. Nucl. Phys. \textbf{15}, 438-450 (1972)
IPTI-381-71.

\bibitem{Altarelli:1977zs}
G.~Altarelli and G.~Parisi,
Nucl. Phys. B \textbf{126}, 298-318 (1977).

\bibitem{COMPASS:2010hwr}
M.~G.~Alekseev \textit{et al.} [COMPASS],
Phys. Lett. B \textbf{693}, 227-235 (2010).

\bibitem{HERMES:2003gbu}
A.~Airapetian \textit{et al.} [HERMES],
Phys. Rev. Lett. \textbf{92}, 012005 (2004).

\bibitem{HERMES:2004zsh}
A.~Airapetian \textit{et al.} [HERMES],
Phys. Rev. D \textbf{71}, 012003 (2005).

\bibitem{Sato:2016tuz}
N.~Sato \textit{et al.} [JAM],
Phys. Rev. D \textbf{93}, no.7, 074005 (2016).

\bibitem{Zhou:2022wzm}
Y.~Zhou \textit{et al.} [JAM],
Phys. Rev. D \textbf{105}, no.7, 074022 (2022).

\bibitem{Deur:2018roz}
A.~Deur, S.~J.~Brodsky and G.~F.~De T\'eramond,
Rept. Prog. Phys. \textbf{82}, 076201 (2019).

\bibitem{Yang:2016plb}
Y.~B.~Yang, R.~S.~Sufian, A.~Alexandru {\it et al.},
Phys. Rev. Lett. \textbf{118}, no.10, 102001 (2017).

\bibitem{Ji:1996ek}
X.~D.~Ji,
Phys. Rev. Lett. \textbf{78}, 610-613 (1997).


\bibitem{Bhattacharya:2022vvo}
S.~Bhattacharya, R.~Boussarie and Y.~Hatta,
Phys. Rev. Lett. \textbf{128}, no.18, 182002 (2022).

\bibitem{Bhattacharya:2017bvs}
S.~Bhattacharya, A.~Metz and J.~Zhou,
Phys. Lett. B \textbf{771}, 396-400 (2017).

\bibitem{Cocuzza:2023oam}
C.~Cocuzza \textit{et al.} [JAM],
Phys. Rev. Lett. \textbf{132}, no.9, 091901 (2024).

\bibitem{Gupta:2018lvp}
R.~Gupta, B.~Yoon, T.~Bhattacharya {\it et al.},
Phys. Rev. D \textbf{98}, no.9, 091501 (2018).

\bibitem{Diehl:2003ny}
M.~Diehl,
Phys. Rept. \textbf{388}, 41-277 (2003).

\bibitem{Burkardt:2002hr}
M.~Burkardt,
Int. J. Mod. Phys. A \textbf{18}, 173-208 (2003).


\bibitem{Mondal:2015uha}
C.~Mondal and D.~Chakrabarti,
Eur. Phys. J. C \textbf{75}, no.6, 261 (2015).

\bibitem{Pasquini:2006dv}
B.~Pasquini and S.~Boffi,
Phys. Rev. D \textbf{73}, 094001 (2006).

\bibitem{Chakrabarti:2013gra}
D.~Chakrabarti and C.~Mondal,
Phys. Rev. D \textbf{88}, no.7, 073006 (2013).

\bibitem{deTeramond:2018ecg}
G.~F.~de Teramond \textit{et al.} [HLFHS],
Phys. Rev. Lett. \textbf{120}, no.18, 182001 (2018).

\bibitem{Lin:2023ezw}
B.~Lin \textit{et al.} [BLFQ],
Phys. Lett. B \textbf{847}, 138305 (2023).

\bibitem{Kaur:2023lun}
S.~Kaur \textit{et al.} [BLFQ],
Phys. Rev. D \textbf{109}, no.1, 014015 (2024).

\bibitem{Liu:2024umn}
Y.~Liu \textit{et al.} [BLFQ],
Phys. Lett. B \textbf{855}, 138809 (2024).

\bibitem{Alexandrou:2020zbe}
C.~Alexandrou, K.~Cichy, M.~Constantinou {\it et al.},
Phys. Rev. Lett. \textbf{125}, no.26, 262001 (2020).

\bibitem{Lin:2020rxa}
H.~W.~Lin,
Phys. Rev. Lett. \textbf{127}, no.18, 182001 (2021).

\bibitem{Bhattacharya:2022aob}
S.~Bhattacharya, K.~Cichy, M.~Constantinou {\it et al.},
Phys. Rev. D \textbf{106}, no.11, 114512 (2022).

\end{thebibliography}

\end{document}